\documentclass[aps,preprint]{revtex4}%
\usepackage{amsfonts}
\usepackage{amsmath}
\usepackage{amssymb}
\usepackage{graphicx}%
\begin{document}
\preprint{ }
\title[ ]{Magnetic Catalysis in the Higher-Order Quark Sigma Model}
\author{M. Abu-Shady}
\affiliation{{\small Department of Applied Mathematics, Faculty of Science, Menoufia
University, Egypt}}
\author{}
\affiliation{}
\author{}
\affiliation{}
\affiliation{}
\pacs{PACS number}

\begin{abstract}
The effect of the higher-order mesonic interactions is investigated on the
chiral symmetry breaking in the presence of an external magnetic field. The
effective of higher-order mesonic potential is employed and is numerically
solved in the mean-field approximation. The chiral symmetry breaking increases
with increasing magnetic field. Two sets of free parameterization are
investigated on the magnetic catalysis. A comparison is discussed with the
original sigma model and other studies. The obtained results are included that
the higher-order mesonic interactions play an important role in the magnetic catalysis.



\end{abstract}
\volumeyear{ }
\volumenumber{ }
\issuenumber{ }
\eid{ }
\startpage{1}
\endpage{ }
\maketitle

\section{Introduction}

Chiral symmetry breaking is an important phenomenon in hadron physics and is
of fundamental importance for hadron properties. The difficulties involved in
obtaining low-energy properties directly from QCD, the fundamental theory of
strong interactions, have motivated the construction of effective models due
to their simplicity and effectiveness in describing hadrons at low energies
[1]. The linear sigma model has been proposed as a model for strong nuclear
interactions [2]. The model was first proposed in the 1960s as a model for
pion-nucleon interactions. Today it serves as an effective model for the
low-energy phase (zero temperature) of quantum chromodynamics [3-6] and its
modification is suggested as Refs. [7-10] to provide a good description of the
baryon properties. In addition, the quark sigma model is successfully applied
to the description of static and dynamic baryon properties at finite
temperature and density as in Refs. [11-15].

The study of the influence of external magnetic fields on the fundamental
properties of quantum chromodynamic (QCD) theory, confinement and dynamical
chiral symmetry breaking is still\ a matter of great interest theoretical and
experimental activities $\left[  16-19\right]  $. In Ref. $\left[  20\right]
$, the chiral symmetry structure in the original sigma model in the presence
of an external uniform magnetic field is investigated. Authors of Ref.
$\left[  21\right]  $ examined the chiral phase transition in the presence of
electromagnetic field and they found the magnetic field enhances the chiral
symmetry breaking in the NJL model. Shushpanov and Smilga $\left[
22,23\right]  $ studied the quark condensate in the presence of external
magnetic field using the Schwinger-Dyson equation. In Refs. $\left[
24-26\right]  $, the proper time method is applied to a four-fermion
interaction model to study the influence of a magnetic field. It is shown that
an external magnetic field has the effect of enhancing chiral symmetry breaking.

Recently, the higher-order mesonic interactions play an important role to test
nonperturbative chiral dynamics $\left[  27\right]  $. On same lines, the
hadron properties are improved in comparison with other models and are in good
agreement with experimental data by including higher-order mesonic
interactions as Refs. $\left[  10,28\right]  $

In this paper, we investigate the effect of higher-order mesonic interactions
on the chiral symmetry breaking in the present of an external magnetic field
in the framework of quark sigma model. In addition, a new parametrization for
sigma mass and coupling constant on the behavior of the phase transition is
studied. So far no attempt has been made to include higher-order mesonic
interactions on the chiral symmetry breaking in the presence of external
magnetic field.

This paper is organized as follows: The original sigma model is briefly
presented in Sec. 2. Next, the effective higher-order mesonic potential in the
presence of external magnetic field is presented in Sec. 3. The results are
discussed and are compared with other models in Secs. 4 and 5, respectively.
Finally, the summary and conclusion are presented in Sec. 6.

\section{The chiral sigma model with original effective potential}

The interactions of quarks via the exchange of $\sigma-$ and $\mathbf{\pi}$-
meson fields are given by the Lagrangian density $[3]$ as follows:
\begin{equation}
{\small L}\left(  r\right)  {\small =i}\overline{\Psi}{\small \partial}_{\mu
}{\small \gamma}^{\mu}{\small \Psi+}\frac{1}{2}\left(  \partial_{\mu}%
\sigma\partial^{\mu}\sigma+\partial_{\mu}\mathbf{\pi}.\partial^{\mu
}\mathbf{\pi}\right)  {\small +g}\overline{\Psi}\left(  \sigma+i\gamma
_{5}\mathbf{\tau}.\mathbf{\pi}\right)  {\small \Psi-U}_{1}{\small (\sigma
,\pi),} \tag{1}%
\end{equation}

with
\begin{equation}
{\small U}_{1}{\small (\sigma,\pi)=}\frac{\lambda^{2}}{4}\left(  \sigma
^{2}+\mathbf{\pi}^{2}-\nu^{2}\right)  ^{2}+{\small m}_{\pi}^{2}{\small f}%
_{\pi}{\small \sigma.}\tag{2}%
\end{equation}
$U_{1}\left(  \sigma,\mathbf{\pi}\right)  $ is the meson-meson interaction
potential where $\Psi,\sigma$ and $\mathbf{\pi}$ are the quark, sigma, and
pion fields, respectively. In the mean-field approximation, the meson fields
are treated as time-independent classical fields. This means that we replace
the power and products of the meson fields by corresponding powers and the
products of their expectation values. The meson-meson interactions in Eq. (2)
lead to hidden chiral $SU(2)\times SU(2)$ symmetry with $\sigma\left(
r\right)  $ taking on a vacuum expectation value \ \ \ \ \ \ \
\begin{equation}
\ \ \ \ {\small \ \ }\left\langle \sigma\right\rangle {\small =-f}_{\pi
}{\small ,}\tag{3}%
\end{equation}
where $f_{\pi}=93$ MeV is the pion decay constant. The final \ term in Eq. (2)
included to break the chiral symmetry explicitly. It leads to the partial
conservation of axial-vector isospin current (PCAC). The parameters
$\lambda^{2}$and $\nu^{2}$ can be expressed in terms of$\ f_{\pi}$, sigma and
pion masses as,
\begin{equation}
{\small \lambda}^{2}{\small =}\frac{m_{\sigma}^{2}-m_{\pi}^{2}}{2f_{\pi}^{2}%
}{\small ,}\tag{4}%
\end{equation}%
\begin{equation}
{\small \nu}^{2}{\small =f}_{\pi}^{2}{\small -}\frac{m_{\pi}^{2}}{\lambda^{2}%
}{\small .}\tag{5}%
\end{equation}

\section{The effective higher-order mesonic potential in the presence of
magnetic field}

In this section, the higher-order mesonic potential $U_{2}(\sigma,\pi)$ is
employed. In Eq. (6), the effective higher-order mesonic is included with the
external magnetic field at zero temperature and density as follows,%

\begin{equation}
U_{eff}(\sigma,\pi)=U_{2}(\sigma,\pi)+U_{Vaccum}+U_{Matter}, \tag{6}%
\end{equation}
where%
\begin{align}
U_{2}\left(  \sigma,\mathbf{\pi}\right)   &  =\frac{\lambda_{1}^{2}}{4}\left(
\sigma^{2}+\mathbf{\pi}^{2}-\nu_{1}^{2}\right)  ^{2}+\frac{\lambda_{2}^{2}}%
{4}\left(  \left(  \sigma^{2}+\mathbf{\pi}^{2}\right)  ^{2}-\nu_{2}%
^{2}\right)  ^{2}\nonumber\\
&  +m_{\pi}^{2}f_{\pi}\sigma\text{.} \tag{7}%
\end{align}
In Eq. 7, the higher-order mesonic potential satisfies the chiral symmetry
when $m_{\pi}\longrightarrow0$ as well as in the standard potential in Eq. 2.
Spontaneous chiral symmetry breaking gives a nonzero vacuum expectation for
$\sigma$ and the explicit chiral symmetry breaking term in Eq. 7 gives the
pion its mass.%
\begin{equation}
\left\langle \sigma\right\rangle =-f_{\pi}. \tag{8}%
\end{equation}
Where%
\begin{equation}
\lambda_{1}^{2}=\frac{1}{4f_{\pi}^{2}}(m_{\sigma}^{2}-m_{\pi}^{2}), \tag{9}%
\end{equation}%
\begin{equation}
\nu_{1}^{2}=f_{\pi}^{2}-\frac{m_{\pi}^{2}}{2\lambda_{1}^{2}}, \tag{10}%
\end{equation}%
\begin{equation}
\lambda_{2}^{2}=\frac{1}{16f_{\pi}^{6}}(m_{\sigma}^{2}-3m_{\pi}^{2}), \tag{11}%
\end{equation}%
\begin{equation}
\nu_{2}^{2}=f_{\pi}^{4}-\frac{m_{\pi}^{2}}{4\lambda_{2}^{2}f_{\pi}^{2}}.
\tag{12}%
\end{equation}
For details, see Refs. $\left[  10,28\right]  $. To include the external
magnetic field in the present model, we follow Ref. $\left[  19\right]  $. So,
the vacuum energy potential is given by
\begin{equation}
{\small U}_{Vaccum}=\frac{N_{c}N_{f}~g^{4}}{(2\pi)^{2}}(\sigma^{2}+\pi
^{2})^{2}(\frac{3}{2}-\ln(\frac{g^{2}(\sigma^{2}+\pi^{2})}{\Lambda^{2}})),
\tag{13}%
\end{equation}
where $N_{c}=3$ and $N_{f}=2$ are color and flavor degrees of freedom.
respectively, and $\Lambda$ is mass scale,
\begin{equation}
U_{Matter}=\frac{N_{c}}{2\pi^{2}}\sum_{f~=u}^{d}(\left\vert q_{f}\right\vert
B)^{2}[\zeta^{(1,0)}(-1,x_{f})-\frac{1}{2}(x_{f}^{2}-x_{f})\ln x_{f}%
+\frac{x_{f}^{2}}{4}] \tag{14}%
\end{equation}
In Eq. 14, we have used $x_{f}=\frac{g^{2}(\sigma^{2}+\pi^{2})}{(2\left\vert
q_{f}\right\vert B)}$ and $\zeta^{(1,0)}(-1,x_{f})=\frac{d~\zeta(z,x_{f})}%
{dz}\mid_{z~=-1}$ that represents the Riemann-Hurwitz function, and also
$\left\vert q_{f}\right\vert $ is the absolute value of quark electric charge
in external magnetic field with intense $B$. In Eq. 6, the of effect of the
finite temperature and chemical potential is not included in the present
model, in which the present model focus on the study of magnetic catalysis at
low energy.

\section{ Discussion of results}

In this section, we study the effect of external magnetic field on the
symmetry breaking in the presence of higher-order mesonic interactions and
also discuss the effect of coupling constant $(g)$ and sigma mass $(m_{\sigma
})$ on symmetry breaking. For this purpose, we numerically calculate the
effective potential in Eq. (6). The parameters of the present model are the
coupling constant $g$ and the sigma mass $m_{\sigma}$. The choice of free
parameters of $g$ and $m_{\sigma}$ based on Ref. $\left[  4\right]  .$ The
parameters are usually chosen so that the chiral symmetry is spontaneously
broken in the vacuum and the expectation values of the meson fields. In this
work, we consider two different sets of parameters in order to get high and a
low value for sigma mass. The first set is given by $\Lambda=16.48$ MeV which
yields $m_{\pi}=138$ MeV and $m_{\sigma}=600$ MeV. The second set as the
first, yielding $m_{\sigma}=400$ MeV.

In Fig. 1, the effect external magnetic field is plotted as a function of
sigma field at $m_{\pi}=0$. Three values of magnetic field $\left(  B\right)
$ are taken. At $B=0$, we note that the spontaneous chiral symmetry breaking
is noted, in which chiral limit is satisfied $(m_{\pi}=0)$. By increasing
magnetic field, we note that the curves shift to upper values and the symmetry
breaking is clearly appeared in the present model. By increasing magnetic
field up to $eB=0.26$~$GeV^{2}$. The potential has minima values. Two minima
values of potential are larger from their values at zero magnetic field. This
indicates that the generation fermion mass increases with strongly increasing
magnetic field. Therefore, the catalysis feature is satisfied in the present
model and agreement with NJL model $\left[  21\right]  $ and Schwinger-Dyson
equation $\left[  22,23\right]  $.%

\begin{center}
\includegraphics[
height=4.0421in,
width=4.0421in
]%
{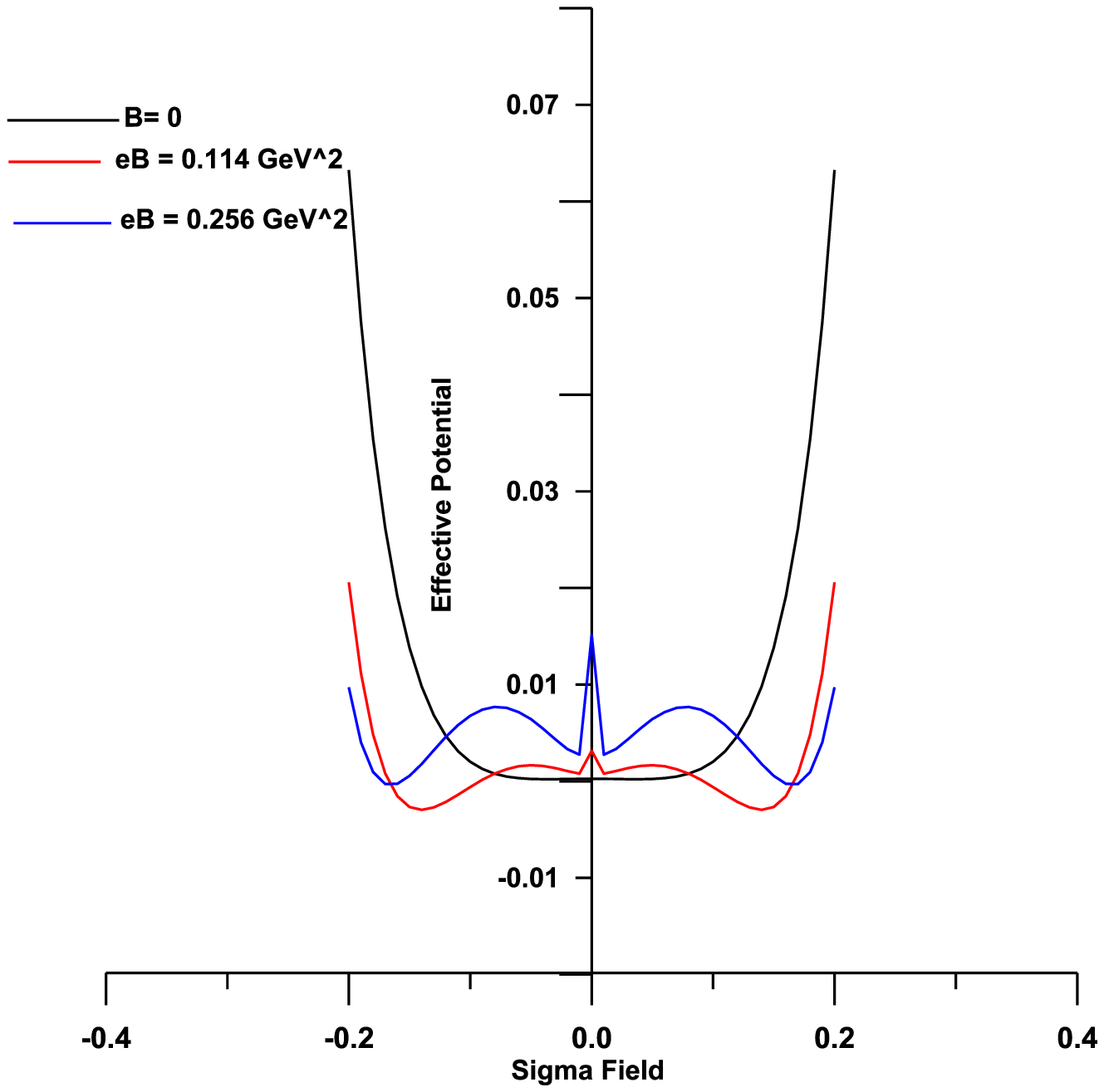}%
\\
{\small Fig. 1: The effective mesonic potential is plotted as a function of
sigma field for different values of magnetic fields }$\left(  B\right)
${\small .}%
\end{center}
%

\begin{center}
\includegraphics[
height=3.5405in,
width=4.0421in
]%
{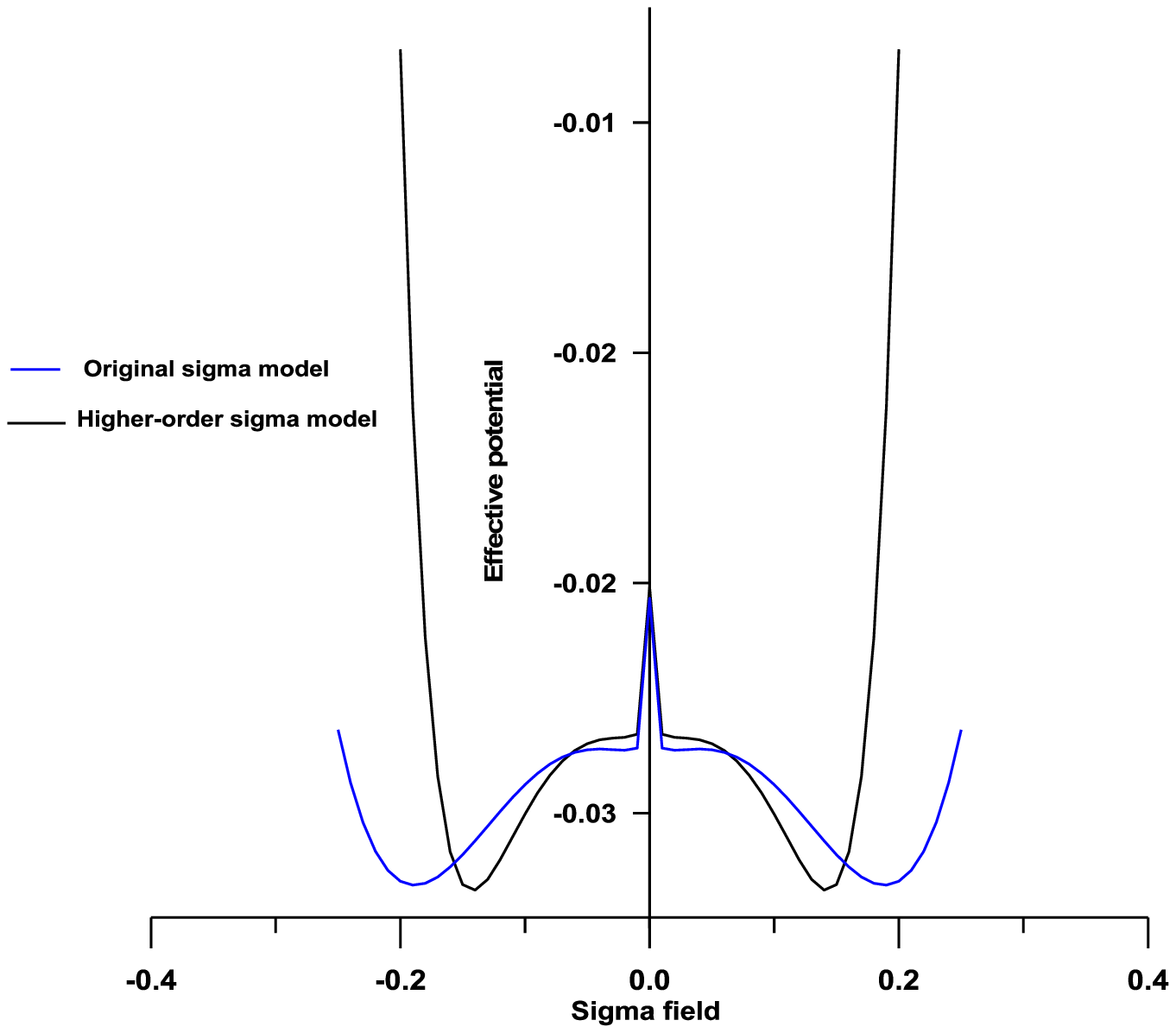}%
\\
{\small Fig. 2: The effective mesonic potential is plotted as a function of
sigma field for the higher-order sigma model and original sigma mode at eB =
0.128 GeV}$^{2}$ {\small and m}$_{\pi}=0${\small .}%
\end{center}

~~\ \ \ \ \ 

In Fig. 2, effective potential of higher-order mesonic interactions is plotted
as a function of sigma field at $eB=0.128$~$GeV^{2}$. We note that qualitative
agreement between the higher-order sigma model and the original sigma model.
The curve of the original model shifts to lower values by including
higher-order mesonic interactions, leading a small change in minima values of
potential which gives generation fermion mass. The effective potential has two
minima values of sigma field in the present model and the original sigma
model. This means that the phase is remain unchanged by including higher-order
mesonic interactions in the present of external magnetic field. This agreement
with the logarithmic sigma model $\left[  28\right]  $.%

\begin{center}
\includegraphics[
height=4.0421in,
width=4.0421in
]%
{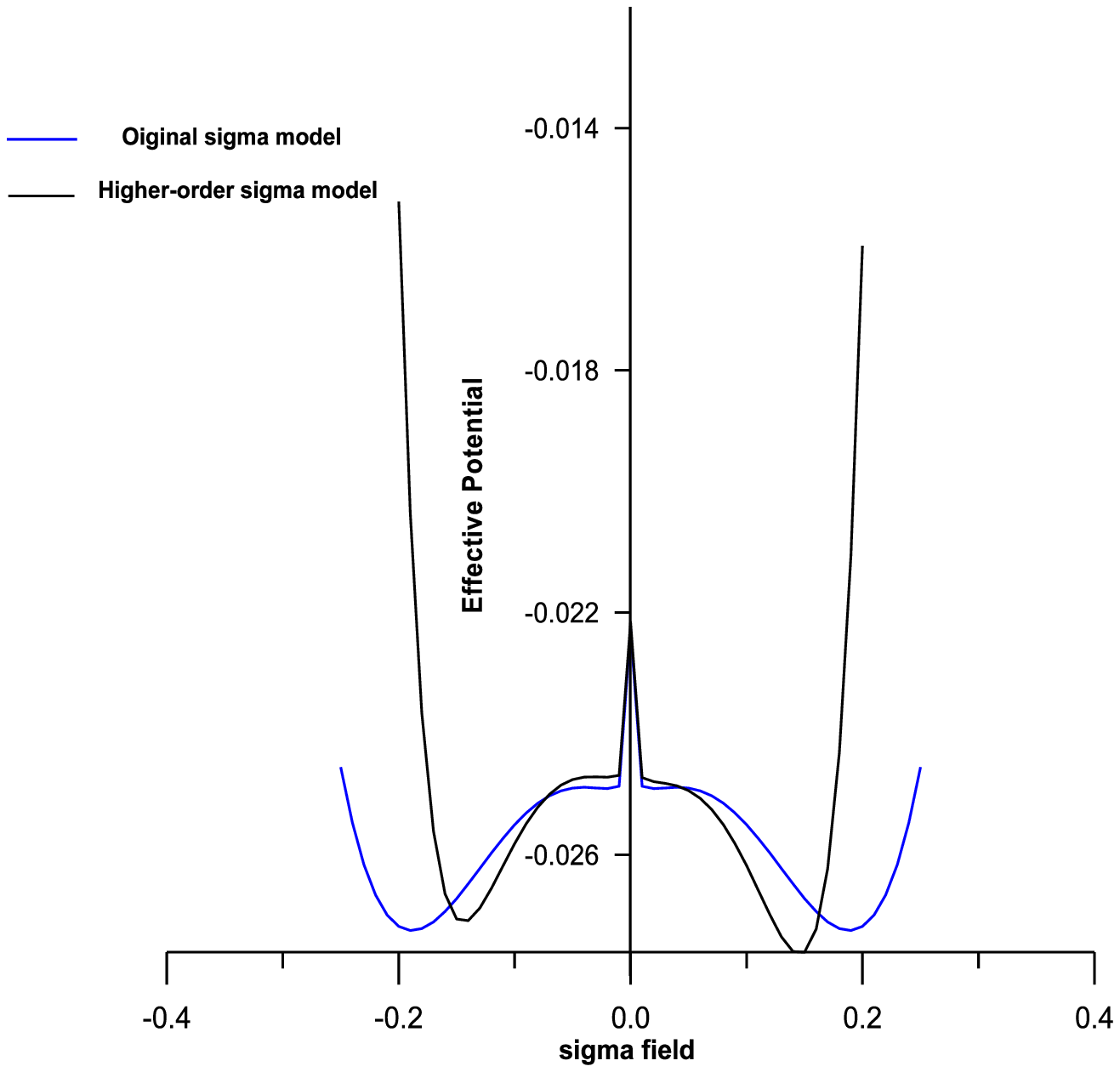}%
\\
{\small Fig. 3: The effective mesonic potential is plotted as a function of
sigma field for the present model and the original sigma model in at eB =
0.128 GeV}$^{2}${\small and m}$_{\pi}=0.14${\small  GeV}%
\end{center}
%

\begin{center}
\includegraphics[
height=4.0421in,
width=4.0421in
]%
{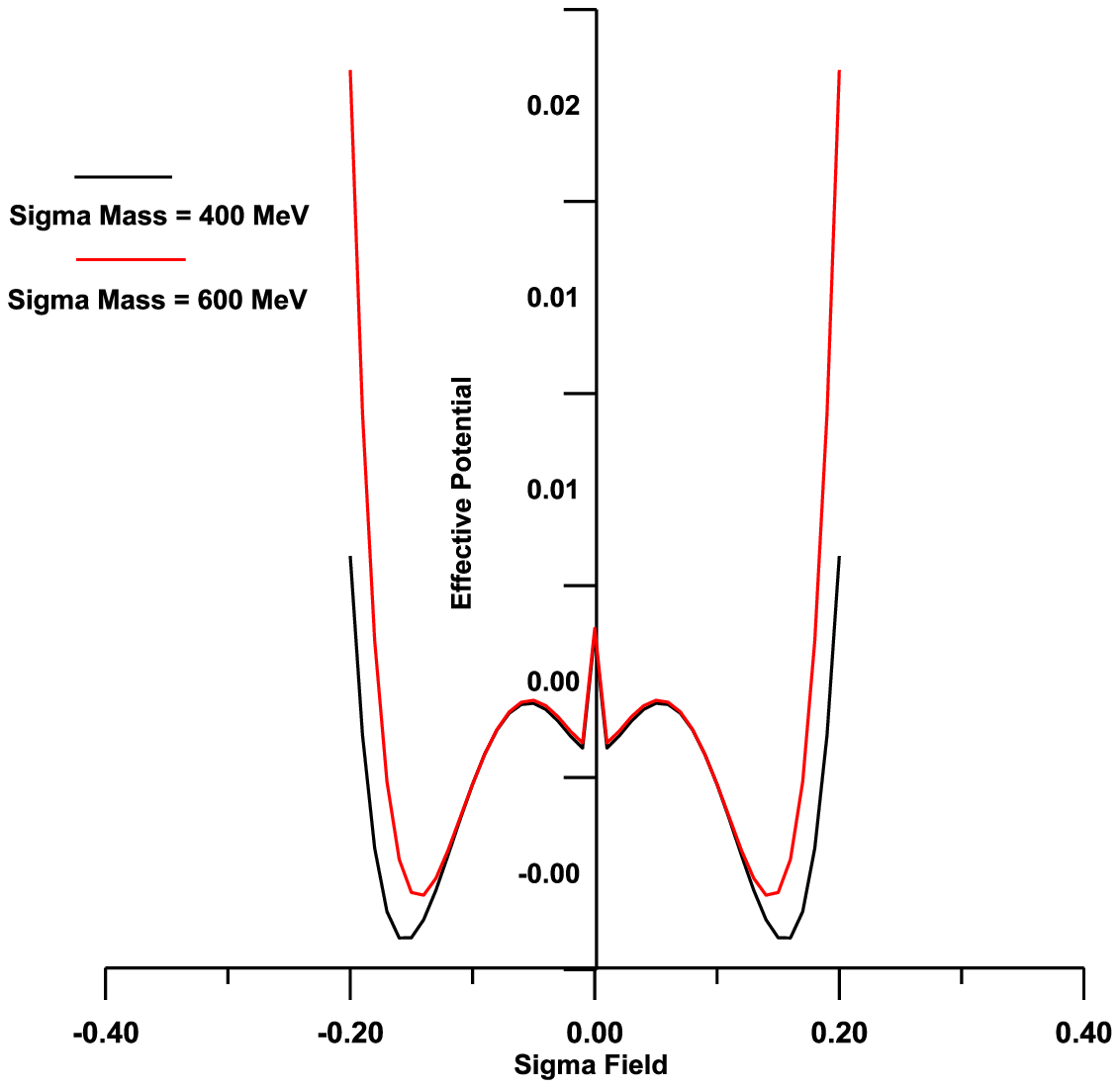}%
\\
{\small Fig. 4: The effective mesonic potential is plotted as a function of
sigma field \ for different values of sigma mass at eB = 0.128 GeV}$^{2}%
${\small and m}$_{\pi}=0${\small .}%
\end{center}
\ \ \ \ \ \ \ \ \ \ 

In Fig. 3, the effective potential is plotted as a function of sigma field
where the explicit symmetry breaking is included $(m_{\pi}=0.14GeV)$ at
$eB=0.128$ GeV$^{2}$. The explicit symmetry breaking is clearly appeared in
the higher-order sigma model in comparison with the original sigma model. \ In
addition, the phase transition is changed from first-order to crossover in the
present model since the pionic mass with magnetic field are included.

It is important to investigate the effect of the role of the free parameters
on the phase transition and symmetry breaking. In Fig. 4, the effective
potential is plotted as a function of sigma field for two values of sigma
mass. We note the curve shifts to lower values by decreasing sigma mass and
the minima values of the potential are larger in comparison their values at
$m_{\sigma}=600$ MeV. Thus, the generated mass by symmetry breaking increases.
In addition, the phase transition is remained unchanged by changing sigma mass
as first-order. In Fig. 5, the effect of coupling constant $\left(  g\right)
$ is drawn. By increasing coupling constant $\left(  g\right)  $, the curve
shifts to lower values with smaller two minima values of potential in
comparison with their values at $g=4.5$. Therefore, the generated mass is
little changed by changing coupling constant $g$ in the present of strong
magnetic field.%

\begin{center}
\includegraphics[
height=4.0421in,
width=4.0421in
]%
{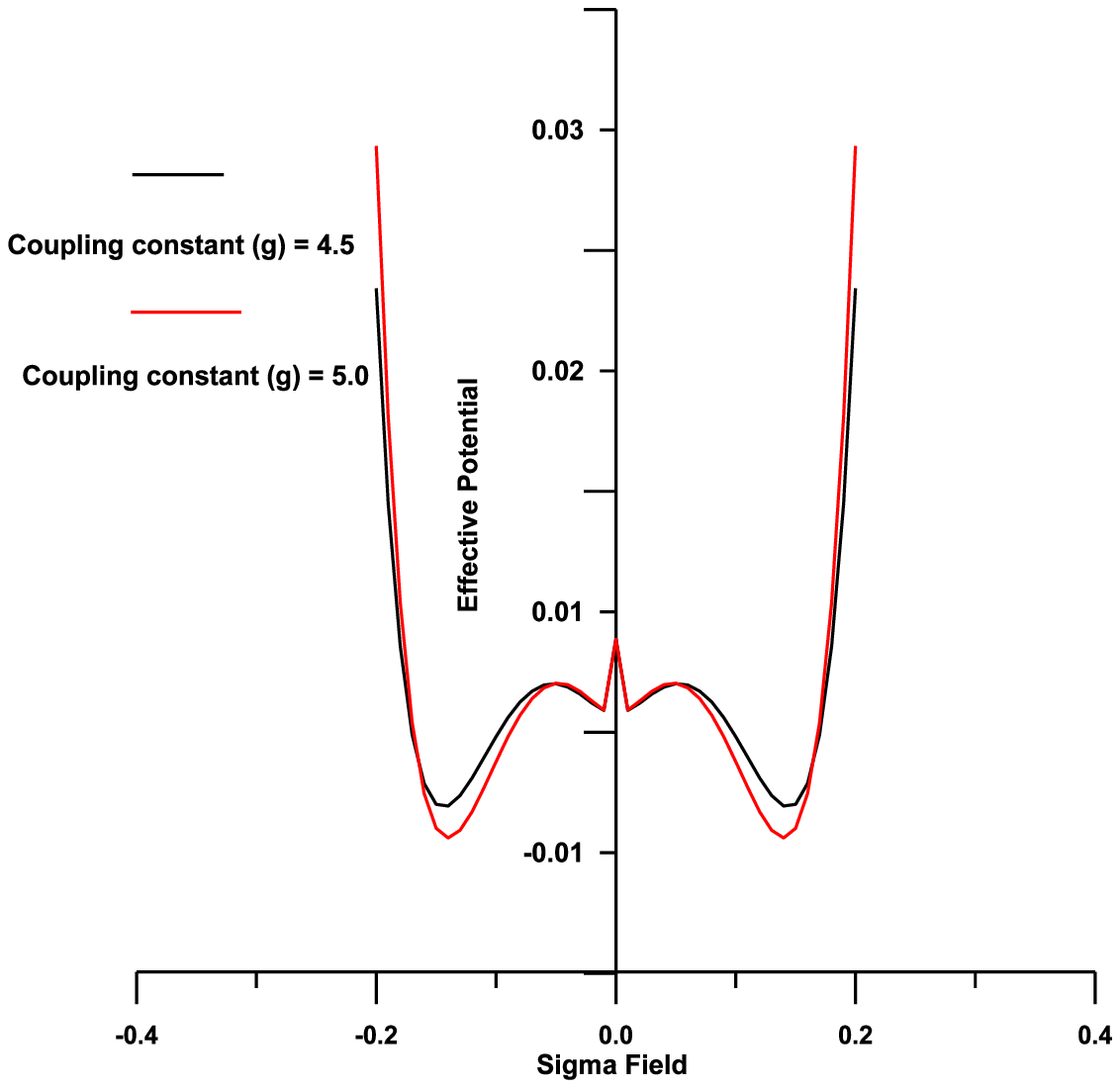}%
\\
{\small Fig. 5: The effective mesonic potential is plotted as a function of
sigma field \ for different values of coupling constant g at eB = 0.128
GeV}$^{2}${\small and m}$_{\pi}=0.$%
\end{center}

\section{Summary and conclusion}

In this work, we have employed the higher-order mesonic potential to study its
effect on the chiral symmetry breaking in the presence of magnetic field. So,
the novelty of this work, that the effect of the magnetic field is not
investigated in the framework of higher-order sigma model. This model applied
only to calculate the hadron properties at low energy without considering
magnetic field as in Refs. $\left[  10,28\right]  $. In addition, the effect
of free parameters of the model is studied in the dense of magnetic field. A
comparison with other models is presented, showing the obtained results are in
the qualitative agreement with\ other models.

Therefore, the chiral higher-order sigma model successfully describes the
magnetic catalysis. In addition, the present study shows that the parameters
of the model play an important role in magnetic catalysis phenomenon. We hope
to extend the present model to finite temperature and chemical potential which
play an important role for studying properties of the universe and neutron star.

\section{\textbf{References}}

\begin{enumerate}
\item V. S. Timoteo and C. L. Lima, Physics Letters B \textbf{635}, 168 (2006).

\item M. Gell-Mann and M. Levy, Nuovo Cinmento \textbf{16}, 705 (1960).

\item M. Birse and M. Banerjee, Phys. Rev. D \textbf{31}, 118 (1985).

\item B. Golli and M. Rosina, Phys. Lett. B \textbf{165}, 347(1985).

\item K. Goeke, M. Harvey, F. Grummer, and J. N. Urbano, Phys. Rev.
D\textbf{37}, 754(1988).

\item T. S. T. Aly, J. A. McNeil, and S. Pruess, Phys. Rev. D \textbf{60},
1114022 (1999).

\item W. Broniowski and B. Golli, Nucl. Phys. A\textbf{\ 714}, 575 (2003).

\item M. Abu-Shady, Int. J. Mod. Phys. A \textbf{26}, 235 (2011).

\item M. Abu-Shady, Int. J. Theor. Phys. \textbf{48}, 1110 (2012).

\item M. Abu-Shady, Phys. Atom. Nuclei \textbf{73}, 944 (2009).

\item N. Bilic and Nikolic, Eur. Phys. J. C \textbf{6}, 515 (1999).

\item O. Scanvenius, A. Moscsy, I. N. Mishustin, and D. H. Rischke, Phys. Rev.
C \textbf{64}, 045202 (2001).

\item H. Mao, T.-Z. Wei and J.-S. Jin, Phys. Rev. C \textbf{88}, 035201 (2013).

\item M. Abu-Shady, Int. J. Mod. Phys. E \textbf{21, }1250061 (2012).

\item M. Abu-Shady, Int. J. Theor. Phys. \textbf{49}, 2425 (2010).

\item D. Kharzeev, K. Landsteiner, A. Schmitt, and H. -U Yee, "Strongly
interacting matter in magnetic fields", Springer, \textbf{624}, 117 (2013).

\item B. S. Kandemir and A. Mogulkoc, Phys. Lett. \textbf{379}, 2120 (2015).

\item K. Kamikado and t. Kanazawa, JHEP \textbf{1}, 129 (2015).

\item G. N. Ferrari, A. F. Garcia, and M. B. Pinto, Phys. Rev. D \textbf{86},
096005 (2012).

\item A. Goyal and M. Dahiya, Phys. Rev. D \textbf{62}, 025022 (2011).

\item S. P. Klevansky and R. H. Lemmar, Phys. Rev. D \textbf{39}, 3478 (1989).

\item I. A. Shushpanov and A. V. Smilga, Phys. Lett. B \textbf{16}, 402 (1997).

\item I. A. Shushpanov and A. V. Smilga, Phys. Lett. B \textbf{16}, 351 (1997).

\item H. Suganuma and T. Tastsumi, Annals. Phys. 208, 470 (1991).

\item K. G. Klimenko and T. Mat. Fiz. \textbf{89}, 211 (1991).

\item V. P. Gusynin, V. A. Miransky, and I. A. Shovkovy, Phy. Rev. Lett.
\textbf{73}, 3499 (1994).

\item A. Feijoo, V. K. Magas, A. Ramos, and E. Oset, Phys. Rev. D \textbf{92},
076015 (2015).

\item M. Abu-Shady and M. Soleiman, Phys. \ Part. and Nucl. Lett. \textbf{10},
683 (2013).

\item M. Abu-Shady, Appl. Math. Inf. Sci. Lett. \textbf{4}, 5 (2016).
\end{enumerate}

\bigskip

\bigskip
\end{document}